\documentclass[article]{emulateapj}

\usepackage{textcomp}
\usepackage{wasysym}
\usepackage{amssymb}
\usepackage{subfigure}
\usepackage{float}
\usepackage{graphicx}
\usepackage{booktabs}

\newcommand{\kph}{km\,s$^{-1}$}
\newcommand{\ddif}{$\sigma_{\rm{CO}} - \sigma_{\rm{HI}}$}
\newcommand{\drat}{$\sigma_{\rm{HI}} / \sigma_{\rm{CO}}$}
\newcommand{\dhi}{$\sigma_{\rm{HI}}$}
\newcommand{\dco}{$\sigma_{\rm{CO}}$}

\newcommand{\cotiiti}{CO $J = 2 \rightarrow 1$}

\newcommand{\hi}{H\,{\sc i}}

\shorttitle{HI and CO Velocity Dispersions}
\shortauthors{Mogotsi et al.}

\begin{document}

\title{HI and CO Velocity Dispersions in Nearby Galaxies}

\author{K.M. Mogotsi \altaffilmark{1}, 
W.J.G. de Blok \altaffilmark{2,1,3},
A. Cald{\'u}-Primo \altaffilmark{4}, 
F. Walter \altaffilmark{4}, 
R. Ianjamasimanana \altaffilmark{5}, 
A.K. Leroy \altaffilmark{6,7}}

\altaffiltext{1}{Astrophysics, Cosmology and Gravity Centre, Department of Astronomy, University of Cape Town, Private Bag X3, Rondebosch 7701, South Africa}
\altaffiltext{2}{Netherlands Institute for Radio Astronomy (ASTRON), Postbus 2, 7990  AA Dwingeloo, the Netherlands}
\altaffiltext{3}{Kapteyn Astronomical Institute, University of
  Groningen, P.O. Box 800, 9700 AV Groningen, the Netherlands}
\altaffiltext{4}{Max-Planck-Institut f{\"u}r Astronomie,
  K{\"o}nigstuhl 17, D-69117, Heidelberg, Germany}
\altaffiltext{5}{College of Graduate Studies, University of South Africa,
P.O. Box 392, UNISA, 0003, South Africa}
\altaffiltext{6}{Department of Astronomy, The Ohio State University, McPherson Laboratory, 140 West 18th Avenue, Columbus OH, 43210-1173, USA}
\altaffiltext{7}{National Radio Astronomy Observatory, 520 Edgemont Road, Charlottesville, VA 22903, USA}

\altaffiltext{1}{email: moses.mog@gmail.com}

\begin{abstract}

We analyze the velocity dispersions of individual \hi\ and CO profiles
in a number of nearby galaxies from the high-resolution HERACLES CO
and THINGS \hi\ surveys.  Focusing on regions with bright CO emission,
we find a CO dispersion value {\dco} $ = 7.3 \pm 1.7$ \kph.  The
corresponding \hi\ dispersion \dhi $ = 11.7 \pm 2.3$ \kph, yielding a
mean dispersion ratio {\drat}\,$= 1.4 \pm 0.2$, independent of radius.
We find that the CO velocity dispersion increases towards lower peak
fluxes. This is consistent with previous work where we showed that
when using spectra averaged (``stacked'') over large areas, larger
values for the CO dispersion are found, and a lower ratio {\drat} $ =
1.0 \pm 0.2$. The stacking method is more sensitive to low-level
diffuse emission, whereas individual profiles trace narrow-line,
GMC-dominated, bright emission.  These results provide further
evidence that disk galaxies contain not only a thin, low velocity
dispersion, high density CO disk that is dominated by GMCs, but also a
fainter, higher dispersion, diffuse disk component.
\end{abstract}

\keywords{galaxies: ISM; ISM: molecules; radio lines: galaxies}

\section{Introduction}

Gas velocity dispersions can be used to estimate the kinetic and
thermal gas temperatures; to determine the mass distribution and
structure of galaxies (e.g., \citealp{petrup07}), and the stability, scale
height and opacity of the gas disk.  Velocity dispersions are
important in studies of star formation, turbulence, the interstellar
medium (ISM), and dynamics of galaxies.  This is especially true of
the vertical velocity dispersion $\sigma_{z}$.  The rotation of the
galactic disk has no effect on this component of the observed
dispersion, and this makes it a useful parameter for studying the
vertical structure of galactic disks.  Dispersions are used to
determine the stability of galactic disks against gravitational
collapse using the Toomre parameter {\citep{toomre64,ken89}}.
Another link to star formation and turbulence studies is that
dispersions can be used to determine the energy of the ISM
(e.g., {\citealp{agtz09,tamb09}}).  They are also important in
determining the midplane pressure of the gas disk
{\citep{elm89,ler08}} and in star formation laws that consider a
variable disk free-fall time {\citep{elm89,krumck05,ler08}}.
{\citet{lars81}} used small-scale internal velocity dispersions to
determine that molecular clouds are dominated by turbulent motions.
Studies at larger scales can be used to determine the level of
turbulence found between giant molecular clouds and in large scale
motions of gas in galaxies.

Studies of velocity dispersions require high spatial and velocity
resolution observations (for vertical velocity dispersion studies,
galaxies of low inclination are required so as to minimize the
contribution from the radial and azimuthal dispersion components).
The effective dispersion ($\sigma_{\rm eff}$) can be thought of as a
combination of the thermal broadening ($v_{t}$) and turbulent
dispersion ($\sigma_{t}$) :
\begin{equation}
\sigma_{\rm eff}^{2} = v_{t}^{2} + \sigma_{t}^{2}
\end{equation}
(e.g., {\citealp{agtz09}}).  The turbulent component can be decomposed
into a radial ($\sigma_{r}$), angular ($\sigma_{\phi}$), and vertical
($\sigma_{z}$) component, or a planar ($\sigma_{xy}$) and vertical
component ($\sigma_{z}$).  Theory and simulations show that the
velocity dispersion is expected to be anisotropic, with $\sigma_{r}
> \sigma_{\phi} > \sigma_{z}$ and $\sigma_{xy}
\sim 2\sigma_{z}$ {\citep{agtz09}}.  When the beam of a telescope is
large compared to the rotational velocity gradient in the observed
galaxy (e.g., in high-redshift galaxies and highly inclined galaxies), beam
smearing can affect the measured dispersion.  For gas components with
a clumpy structure (e.g., molecular gas), there are additional
complications: the observed dispersion $\sigma_{\rm obs}$ is then a combination of the
dispersion between clouds (cloud-cloud dispersions $\sigma_{\rm c-c}$)
and the internal velocity dispersion within the clouds
($\sigma_{\rm internal}$):

\begin{equation}
\sigma_{obs}^{2} = \sigma_{c-c}^{2} + \sigma_{\rm internal}^{2}.
\end{equation}

The structure of {\hi} is more filamentary and less clumpy than that
of molecular gas. Its velocity dispersion is therefore generally not
decomposed into internal and cloud-cloud components.  

\subsection{{\rm \hi} velocity dispersions}

Since {\hi} is the dominant gas component of galaxies and is easily
observable through the $21$\,cm emission line, it has been extensively
studied.  {\hi} velocity dispersions of nearby galaxies have been 
well studied, most notably by {\citet{petrup07}} and {\citet{tamb09}}.
Early work showed that {\dhi} $\sim 6-13$\,\kph\ (e.g.,
{\citealt{shovnk84}}, {\citealt{vnksho84}}, {\citealt{kmpsanc93}}),
with the dispersions dropping with increasing radial distance from the
center (e.g., {\citealt{kmpsanc93}}).  {\citet{hunter01}} and {\citet{hunter11}} also studied {\dhi} in dwarf galaxies.  {\citet{petrup07}} performed
high-resolution and high-sensitivity {\hi} observations of the nearly
face-on galaxy NGC 1058 to study its gas velocity dispersion.  They
found a vertical velocity dispersion of $4-14$\,\kph, which decreased
with radius. These studies reached resolutions of $\sim
600$\,pc. \citet{tamb09} used high-resolution {\hi} data from 
The {\hi} Nearby Galaxies Survey (THINGS; {\citealt{walt08}}) to study {\hi}
velocity dispersions. They also found that the dispersions decreased
with radius.  They found a mean {\dhi} of $\sim 10$\,\kph\ at
$r_{25}$, dropping off to $\sim 5 \pm 2$\,\kph\ at larger radii.
Stacking analysis was used by {\citet{ianja12}} to study the velocity
dispersions averaged over \hi\ disks of the THINGS galaxies.  They
found {\dhi} $= 12.5 \pm 3.5$\,\kph\ ({\dhi} $= 10.9 \pm
2.1$\,\kph\ for galaxies with inclinations less than 60$^{\circ}$).
This stacking analysis allowed them to study the {\hi} velocity
profiles at high signal-to-noise, enabling them to decompose the {\hi}
profiles into broad and narrow components.  Fitting these components
with Gaussians, they found {\dhi} $= 6.5 \pm 1.5$\,\kph\ for the
narrow (cold) {\hi} component and {\dhi} $= 16.8 \pm 4.3$\,\kph\ for
the broad (warm) {\hi} component. A similar analysis by \citet{stilp13}
of partially the same data found velocity dispersions of the bulk of
the {\hi} of $\sim 6-10$ \kph.

\subsection{CO velocity dispersions}

CO velocity dispersions have been less studied than those of
\hi. Mostly this has been due to technical limitations.  Early
observations of the lowest three CO rotational transitions found
 dispersions in the range of $5 - 9$
\kph\ \citep{stark84,wilsco90,combec97,walsh02,wilson11}.  Recent
instrumental developments have enabled more extensive studies of the
CO distribution in galaxies, such as the HERA CO Line
Extragalactic Survey (HERACLES; {\citealt{ler09}}); see also Section
{\ref{sec:datamethod}}.  HERACLES is a \cotiiti\ survey of nearby
galaxies, covering their entire star-forming disks. It partially
overlaps with the THINGS survey, meaning \hi\ and CO data are
available at comparable resolutions.

\citet{cal13} used data from HERACLES and THINGS to compare CO and
{\hi} velocity dispersions as averaged over large areas using the
stacking technique. They analyzed the dispersions of these stacked
{\hi} and CO velocity profiles, stacking by galactocentric radius,
star formation, {\hi}, CO and total gas density.  They found that
{\dhi}\,$=11.9 \pm 3.1$\,\kph, {\dco}\,$=12.0 \pm 3.9$\,\kph\ with
{\drat}\,$= 1.0 \pm 0.2$. In other words, the CO dispersions they
found are very similar to the {\hi} dispersions.  \citet{cal13}
suggested that this indicates the presence of an additional, more
diffuse, higher dispersion molecular disk component that is similar in
thickness to the {\hi} disk (see also {\citealt{shet14}}).  This finding is in agreement with
independent studies by, e.g., {\citet{gar92}} who find, in addition to
a thin molecular disk, a 2--3 kpc thick molecular ``halo'' around the
edge-on galaxy NGC 891. A similar thick 
molecular disk is also found by \citet{com12} in M33.  {\citet{pet13}} compared interferometric and
single-dish observations of M51 (NGC 5194) and also found evidence
of an extended molecular disk. Similar results have been found by
{\citet{cal15}}, again by comparing interferometric and single-dish
imaging of the molecular gas disks in nearby galaxies.  

The results presented in \citet{cal13} were based on stacked profiles,
i.e., profiles averaged over large regions. In this paper we use the
same THINGS and HERACLES data { as used by \citet{cal13}} to determine whether evidence for the
diffuse molecular component can also be found in individual profiles.
In particular, we investigate whether the velocity dispersion of the
CO profiles changes as a function of CO intensity, which is what
one would expect if a diffuse, high-velocity dispersion component is
indeed present.

In Section {\ref{sec:datamethod}} we describe the data used.  Section
{\ref{sec:results}} contains a description of the results of our
analysis.  Section {\ref{sec:discussion}} contains a discussion of our
results and a comparison to other work. In Section 5 we summarize our conclusions.

\section{Data and method}
\label{sec:datamethod}

We used Hanning-smoothed CO data cubes from HERACLES \citep{ler09},
which is a molecular gas survey of nearby galaxies using the HERA
receiver array on the IRAM 30-m telescope.\footnote[1]{IRAM is
  supported by CNRS/INSU (France), the MPG (Germany), and the IGN
  (Spain)}  For the neutral hydrogen, we used residual-scaled
natural-weighted {\hi} data cubes from THINGS \citep{walt08}, which is a 21-cm survey of 34 nearby
spiral and dwarf galaxies.  The observations were done with the
NRAO\footnote[2]{The National Radio Astronomy Observatory is a
  facility of the National Science Foundation operated under
  cooperative agreement by Associated Universities, Inc.} Jansky Very
Large Array.  The work in this paper is based on the analysis done on
13 galaxies (see Table \ref{tab:props}) that are common to both
surveys and which have CO detections.  The properties of these
galaxies can be found in Table 1 of \citet{walt08}.  For convenience,
noise values and velocity resolutions of the \hi\ and CO observations are listed
in Table \ref{tab:props}. { Note that these are the same data as used in the analysis presented in \citet{cal13}.}

\begin{table}
\caption{Noise and Velocity Resolution of the {\hi} and CO cubes.\label{tab:props}}
\begin{center}
\begin{tabular}{ccccc}
\tableline
\tableline
Galaxy & {\hi} Noise & CO Noise  & {\hi} $\Delta V$  & CO $\Delta V$  \\
 &  [mJy\,beam$^{-1}$] & [mK] & [km\,s$^{-1}$] & [km\,s$^{-1}$] \\
(1) & (2) & (3) & (4) & (5) \\
\tableline
NGC 628 & $0.60$ & $21$ & $2.6$ & $5.2$ \\
NGC 925  & $0.57$ & $16$ & $2.6$ & $5.2$\\
NGC 2403 & $0.38$ & $19$ & $5.2$ & $5.2$\\
NGC 2841 & $0.35$ & $16$ & $5.2$ & $5.2$\\
NGC 2903 & $0.41$ & $21$ & $5.2$ & $5.2$\\
NGC 2976 & $0.36$ & $20$ & $5.2$ & $5.2$\\
NGC 3184 & $0.36$ & $17$ & $2.6$ & $5.2$\\
NGC 3198 & $0.33$ & $17$ & $5.2$ & $5.2$\\
NGC 3351 & $0.35$ & $19$ & $5.2$ & $5.2$\\
NGC 4214 & $0.69$ & $19$ & $1.3$ & $5.2$\\
NGC 4736 & $0.33$ & $21$ & $5.2$ & $5.2$\\
NGC 5055 & $0.36$ & $26$ & $5.2$ & $5.2$\\
NGC 6946 & $0.55$ & $25$ & $2.6$ & $5.2$\\
\tableline
\end{tabular}
\tablecomments{
Column 1: Galaxy name;
Column 2: Noise per channel in {\hi} data;
Column 3: Noise per channel in CO data;
Column 4: {\hi} velocity resolution;
Column 5: CO velocity resolution.}
\end{center}
\end{table}

\begin{figure*}[htp]
\centering
\includegraphics[width=0.8\textwidth]{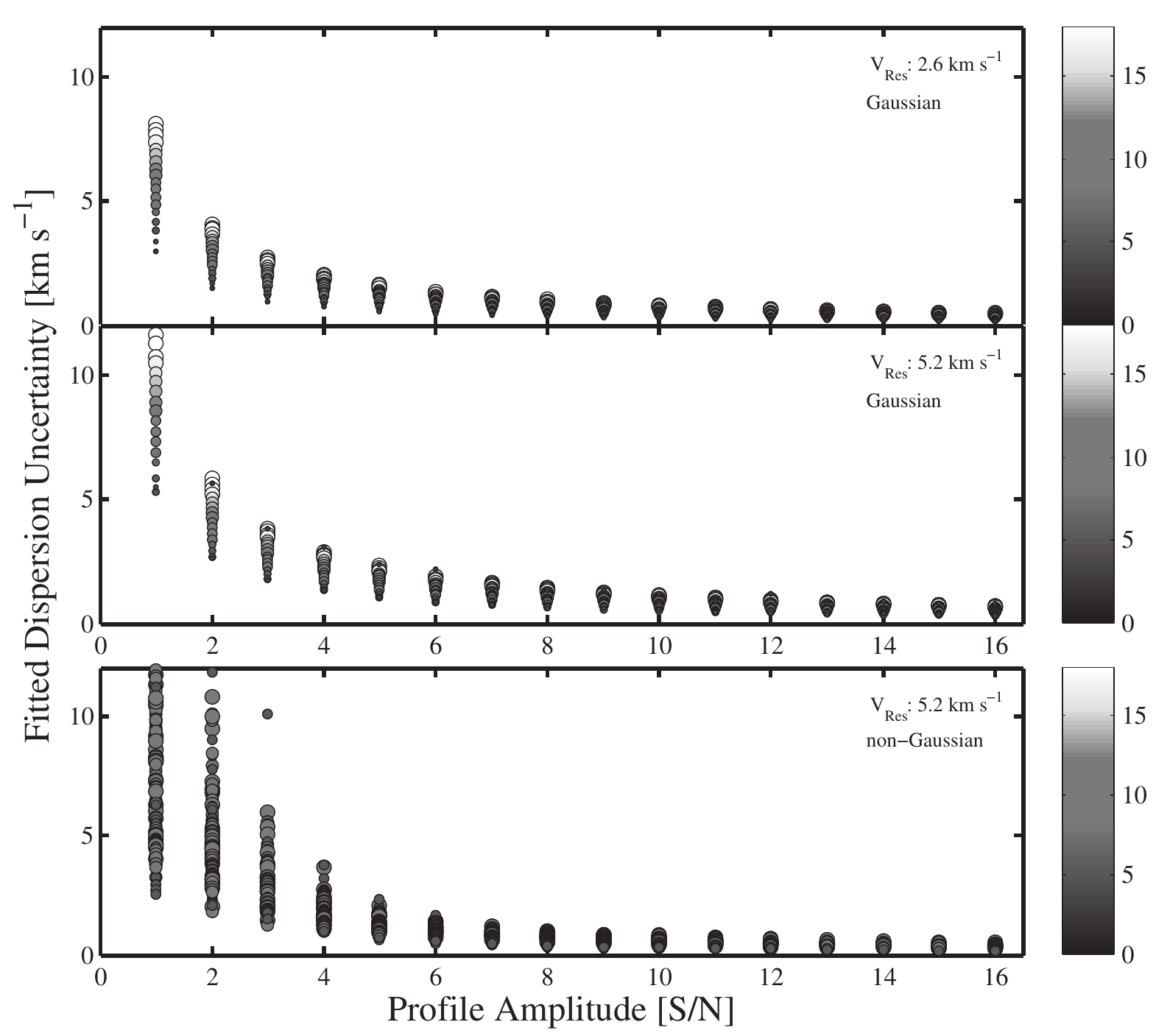}
\caption{Uncertainties in fitted
  dispersions (y-axis) for Gaussian profiles of various amplitudes and
  dispersions.  The input Gaussian amplitudes (in signal-to-noise units)
  are shown on the x-axis.  Marker sizes and grayscale represent input
  velocity dispersions.  {\itshape{Top and middle panels:}} Gaussian velocity profiles were added to random
  Gaussian noise and Gaussians were fitted to the resultant velocity
  profiles.  The mean of 1000
  iterations is plotted for each input amplitude and dispersion value.
  The y-axis values are the mean uncertainties of the fitted
  dispersions. The top panel is for simulated data with a
  velocity resolution of $2.6$ \kph; the middle panel is for a
  resolution of $5.2$ \kph.  {\itshape{Bottom panel:}} Gaussian velocity profiles were added to real noise
  extracted from $5.2$ \kph\ resolution CO data cubes and Gaussians
  were fitted to the resultant velocity profiles.  The noise from the
  cubes was selected from regions with no galactic emission.}
\label{fig:noisesim2}
\label{fig:noisesim}
\end{figure*}

The CO data generally have a spatial resolution of $13''$. 
The natural-weighted {\hi} data cubes mostly have resolutions
better than this, but were smoothed to $13''$ to match the
resolution of the CO data.  

A number of recent studies of {\hi} velocity profiles have used
Gauss-Hermite profiles to take into account asymmetries in the
profiles or used multiple Gaussian components to quantify the presence
of different components of the ISM (e.g., \citealt{deB08},
\citealt{ianja12}). We did explore these fitting functions for our
profiles, but found that the CO profiles are better described by
simple, single Gaussians.  In order to minimize the number of fit
parameters, and as we are only interested in the general width of the
profile, we therefore use single Gaussians to fit both the {\hi} and CO
profiles.

In fitting the profiles we imposed a $4S$ noise cutoff on the fitted
peak fluxes of the profiles, where $S$ is the rms noise of the
profile.\footnote[3]{To avoid confusion with the velocity dispersion
  $\sigma$, we use $S$ throughout this paper to indicate the rms noise
  level.} Only positions where both {\hi} and CO profiles had peak
fluxes greater than $4S$ were retained.  For the {\hi} data,
determining the $4S$ values was done using non-residual-scaled cubes,
as residual-scaling affects the relation between signal and noise (see
\citealt{walt08} for a full description of the residual scaling
procedure).  These results were then applied as a mask to the
residual-scaled cubes.  The remaining velocity profiles of these
masked residual-scaled cubes were then fitted and analyzed. In
addition to the peak-flux criterion, we also imposed a
velocity resolution cutoff where all profiles with fitted dispersions
smaller than the velocity resolution of their data cube were
removed.

We simulated how the uncertainties in the fitted dispersions
behave by producing random Gaussian noise at velocity resolutions
relevant to our data, adding pre-determined Gaussian velocity profiles
to them and re-fitting the data.  Simulations were performed with
input Gaussian profiles of different amplitudes and dispersions.  A
thousand iterations of data simulation and fitting were performed for
each input amplitude and dispersion value.  This was done for
different velocity resolutions and the averages of the fit
uncertainties are plotted in Fig.\ \ref{fig:noisesim} (top and middle
panel).  Input dispersions ranged between $2.6$ \kph\ and $20$ \kph.

For velocity profiles with peak fluxes greater than $4S$ and velocity
resolutions of $2.6$ \kph, the mean uncertainties in the fitted
dispersion were smaller than $\sim 2$~\kph.  For velocity profiles
with peak flux equal to $4S$ and velocity resolutions of $5.2$~\kph,
the mean errors in the fitted dispersion were between $1.4$~\kph\ and
$\sim 3$~\kph.

A small number of the CO spectra showed some minor baseline ripples
resulting in a slightly non-Gaussian noise behaviour.  We therefore
repeated the same procedure but added noise extracted from these CO
cubes rather than random Gaussian noise.  This was done for profiles
with peak values between $1S$ and $16S$.  The results are plotted
in Fig.\ \ref{fig:noisesim} (bottom panel).  The results for the CO
noise simulation are consistent with the results from the Gaussian
noise simulation down to $4S$ peak flux levels.

\section{Results}
\label{sec:results}

\subsection{Comparing {\rm \hi} and CO velocity dispersions}
\label{subsec:pixpixan}

Using the results from the Gaussian fits, \hi\ and CO dispersion maps
were made for each galaxy.  In addition, we made dispersion difference
({\ddif}) and dispersion ratio ({\drat}) maps for each galaxy by
taking the CO and {\hi} dispersion maps and then doing a
pixel-by-pixel subtraction or division.

\begin{figure*}[htp]
\centering
\includegraphics[width=0.8\textwidth]{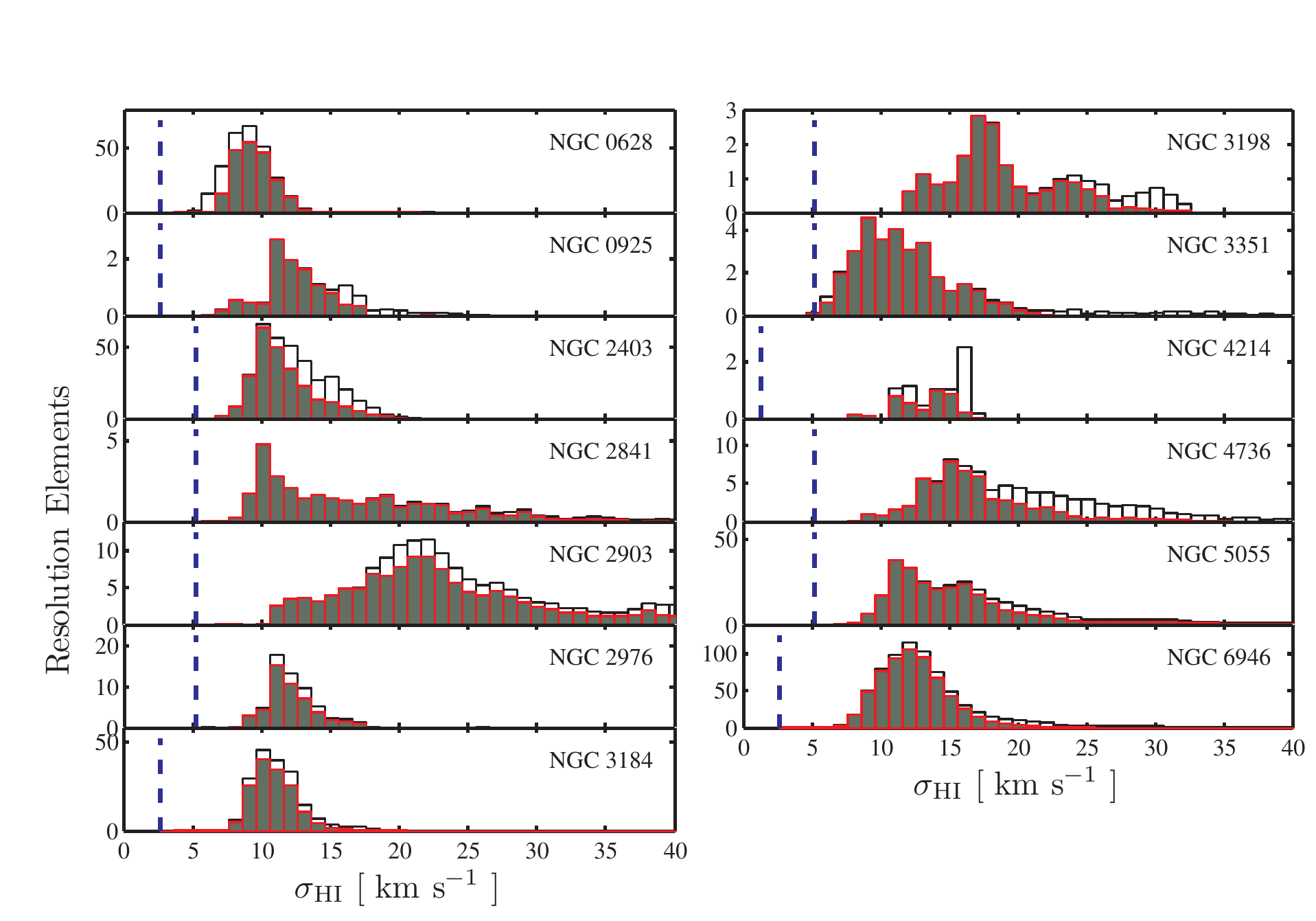}
\caption{Distributions of the \hi\ dispersions {\dhi} for positions
  that have both CO and \hi\ emission.  The velocity dispersion
  cutoffs are indicated by the vertical dotted lines.  The y-axis is
  in resolution elements (resolution elements $=$ [number of
    pixels]/[number of pixels per single resolution element],
  i.e. number of beams).  Open bars show the distribution for all
  profiles. The gray shaded bars with red outlines show the
  distribution of dispersions for profiles with a fit uncertainty less
  than $1.5$ \kph, and outside the central $0.2$\,$r_{25}$ radial
  range.
\label{fig:HIDisperr}
\label{fig:HIDisp}}
\end{figure*}

\begin{figure*}[htp]
\centering
\includegraphics[width=0.8\textwidth]{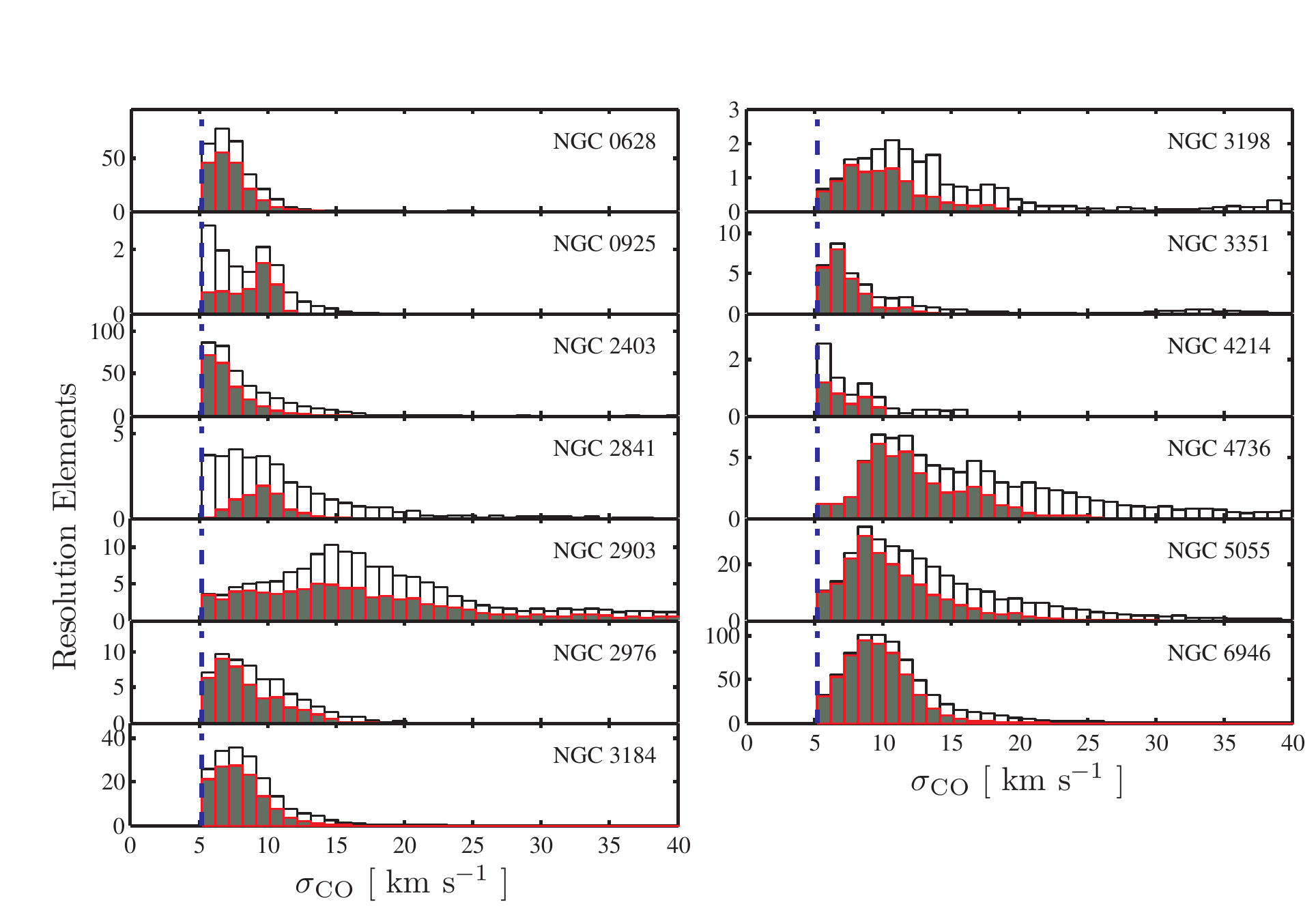}
\caption{Same as Figure {\ref{fig:HIDisp}}, but now for CO dispersions.}
\label{fig:CODisperr}
\label{fig:CODisp}
\end{figure*}

The dispersions were binned into $1$ \kph\ bins. Histograms of the
$\sigma_{\rm{HI}}$ and $\sigma_{\rm{CO}}$ distributions for those
positions in each galaxy where {\hi} and CO were both present are
shown in Figs.\ \ref{fig:HIDisp} and \ref{fig:CODisp}.  The
distribution of dispersion values from pixels outside the central
0.2$r_{25}$ and with small fit uncertainties ($\Delta \sigma \leq 1.5$
\kph) are shown as the shaded histograms in the figures. The
0.2$r_{25}$ selection was used in order to minimize the effect of beam
smearing, as discussed later in this section.  Dispersions are plotted
against the number of resolution elements (defined as the ratio of the
number of pixels and the number of pixels per beam), or, equivalently,
the number of beams.  From the histograms it is clear that in regions
where there is both \hi\ and CO emission, $\sigma_{\rm{HI}}$ values
range from $\sim 5 - 30$ \kph\ and $\sigma_{\rm{CO}}$ values range
from $\sim 5 - 25$ \kph.  The $\sigma_{\rm{HI}}$ modes range from $9-
22$ \kph\ and $\sigma_{\rm{CO}}$ modes range from $6 - 15$ \kph\ (see
Table \ref{tab:modes}).  Most of the high dispersions have large
fitting errors and/or are from pixels in the central regions of
galaxies, as shown in Figs.\ \ref{fig:HIDisp} and \ref{fig:CODisp}.
Such large dispersions are usually due to multiple gas components in
the line of sight and/or beam smearing.  These give non-Gaussian
profiles resulting in bad fits.

\begin{table*}
\begin{center}
\caption{The statistical properties of the \hi\ and CO dispersions.\label{tab:modes}\label{tab:means}}
\begin{tabular}{rccccccc}
\tableline\tableline
 Galaxy & {\dco} & {\dhi} &  \multicolumn{1}{c}{{\dhi} (all HI)} & \multicolumn{2}{c}{\ddif}&\multicolumn{2}{c}{\drat}\\
 & (km\,s$^{-1}$) & (km\,s$^{-1}$) & (km\,s$^{-1}$) & (km\,s$^{-1}$) & (km\,s$^{-1}$) & & \\
(1) & (2) & (3) & (4) & (5) & (6) & (7) & (8) \\
\tableline
NGC 628  & 6 (7.5)  & 9 (9.1) & 6 (6.5) & $-$1.70 ($\pm$0.04)  & $-$2 ($-$1.6) & 1.2 ($\pm$0.01)& 1.2 (1.3) \\
NGC 925  & 6 (8.5)  & 11 (13.2) & 10 (10.9) & $-$3.5 ($\pm$0.3)  & $-$3 ($-$4.7) & 1.4 ($\pm$0.04)& 1.3 (1.7) \\
NGC 2403 & 6 (8.3)  & 10 (12.3) & 7 (8.7) & $-$3.8 ($\pm$0.1) & $-$4 ($-$4.0) & 1.5 ($\pm$0.02)& 1.4 (1.6) \\
NGC 2841 & 8 (10.9)  & 10 (18.0) & 9 (15.5) & $-$3.5 ($\pm$0.9)  & $-$3 ($-$7.1) & 1.3 ($\pm$0.06)& 1.3 (2.0) \\
NGC 2903 & 15 (21.6) & 22 (25.3) & 8 (12.1) & $-$5.5 ($\pm$0.2) & $-$5 ($-$3.7) & 1.3 ($\pm$0.02)& 1.2 (1.4) \\
NGC 2976 & 6 (9.3)  & 11 (12.2) & 11 (13.2) & $-$3.1 ($\pm$0.2) & $-$5 ($-$2.9) & 1.3 ($\pm$0.05)& 1.1 (1.4) \\
NGC 3184 & 8 (8.6) & 10 (11.1) & 7 (8.3) & $-$2.57 ($\pm$0.03) & $-$2 ($-$2.5) & 1.3 ($\pm$0.02)& 1.2 (1.4) \\
NGC 3198 & 11 (15.0) & 17 (20.2) & 11 (12.5) & $-$6.6 ($\pm$0.4)  & $-$5 ($-$5.2) & 1.4 ($\pm$0.06)& 1.2 (1.6) \\
NGC 3351 & 7 (14.1) & 9 (18.9) & 7 (9.8) & $-$2.8 ($\pm$0.3)  & $-$3 ($-$4.7) & 1.4 ($\pm$0.04)& 1.2 (1.5) \\
NGC 4214 & 6 (8.0)  & 16 (14.0) & 6 (7.4) & $-$6.0 ($\pm$0.2) & $-$6 ($-$5.8) & 1.4  ($\pm$0.04)& 1.6 (1.9) \\
NGC 4736 & 10 (17.7)  & 15 (23.5) & 7 (10.4) & $-$3.4 ($\pm$0.1) & $-$4 ($-$3.2) & 1.2 ($\pm$0.02)& 1.0 (1.2) \\
NGC 5055 & 9 (14.9)  & 11 (17.9) & 7 (9.9) & $-$2.9 ($\pm$0.1) & $-$3 ($-$3.0) & 1.2 ($\pm$0.01)& 1.3 (1.3) \\
NGC 6946 & 9 (11.6)  & 13.0 (13.5) & 7 (8.4) & $-$2.4 ($\pm$0.1) & $-$2 ($-$2.0) & 1.2 ($\pm$0.02)& 1.2 (1.3) \\
\tableline
\end{tabular}
\tablecomments{
These values were calculated for all pixels where both the {\hi} and CO are above the noise cutoff, except for column 4 which was calculated for all pixels where the {\hi} was above the noise cutoff.
Column 1: Galaxy name; 
Column 2: Mode (mean) of CO dispersions; 
Column 3: Mode (mean) of \hi\ dispersions; 
Column 4: Mode (mean) of \hi\ dispersions of the entire \hi\ disk; 
Column 5: Gaussian fitted mean (fit uncertainty) of {\ddif}\,; 
Column 6: Mode (mean) of {\ddif};
Column 7: Gaussian fitted mean (fit uncertainty) of {\drat}\,;
Column 8: Mode (mean) of {\drat}.
}
\end{center}
\end{table*}

The $\sigma_{\rm{HI}}$ distributions clearly peak at values much
larger than the dispersion cutoffs imposed due to the velocity
resolution of the data.  However, many of the $\sigma_{\rm{CO}}$
distributions have peaks near the dispersion cutoffs. In a few cases
clear $\sigma_{\rm{CO}}$ distribution peaks are not seen (e.g., NGC
2403), and therefore the true mean (and mode) $\sigma_{\rm{CO}}$
values for these galaxies are likely to be smaller than $5.2$ km
s$^{-1}$.

The incomplete sampling and asymmetry of the velocity dispersion
histograms means that, especially for the CO, the mean is not a good
statistic to characterize the distribution (it will overestimate the
typical dispersion value). We therefore also use the mode to describe
the dispersion distributions.  The modes were calculated after binning
the dispersions using a $1$ \kph\ bin size.  The values are listed in
Table~\ref{tab:modes}.

Most of the {\dco} modes range from $6$ to $11$ \kph\ (12/13 galaxies)
while their means range from $7$ to $15$ \kph\ (11/13 galaxies); most
of the {\dhi} modes range from $9$ to $17$ \kph\ (12/13 galaxies) and
their means range from $9$ to $21$ \kph\ (11/13 galaxies).  NGC 2841,
NGC 2903, NGC 3198 and NGC 3351 were not included in the determination
of the average values due to their high inclinations and very
asymmetric dispersion distributions.  The average {\dco} mode is $7.3
\pm 1.7$ \kph\ (average of the {\dco} means is $10.5 \pm 3.6$ \kph),
and the average {\dhi} mode is $11.7 \pm 2.3$ \kph\ (average of the
{\dhi} means is $14.1 \pm 4.3$\,\kph).  Characteristic values of
{\dhi} and {\dco} are listed in Table~\ref{tab:dispval}.  We also list
the median values there for comparison with results from \citet{cal13}
{ which were derived from stacking the same data we use} (see
Section \ref{sec:discussion}).

\begin{figure}[htp]
\centering
\includegraphics[width=0.45\textwidth]{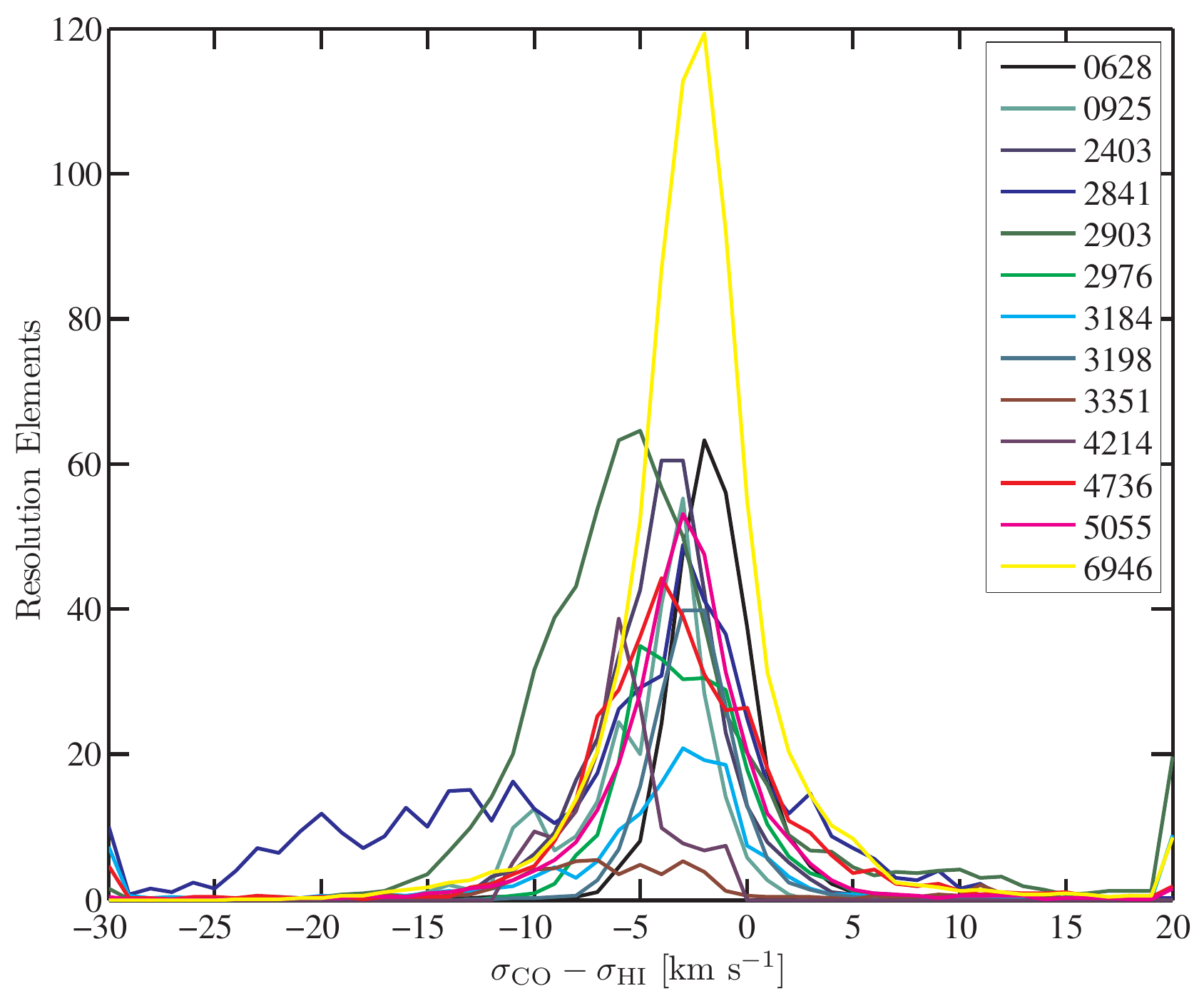}
\caption{Distributions of the dispersion difference {\ddif} values of
  the galaxies.  The lines are colour-coded by galaxy (see legend).
  The y-axis is in resolution elements (resolution elements = [number
    of pixels]/[number of pixels per single resolution
    element]). Values for NGC 925, NGC 2841, NGC 2903, NGC 3198, NGC
  3351, NGC 4214 and NGC 4736 are multiplied by a factor of 5 for
  better comparison with the other galaxies.}
\label{fig:DispDiff}
\end{figure}

\begin{figure}[htp]
\centering
\includegraphics[width=0.45\textwidth]{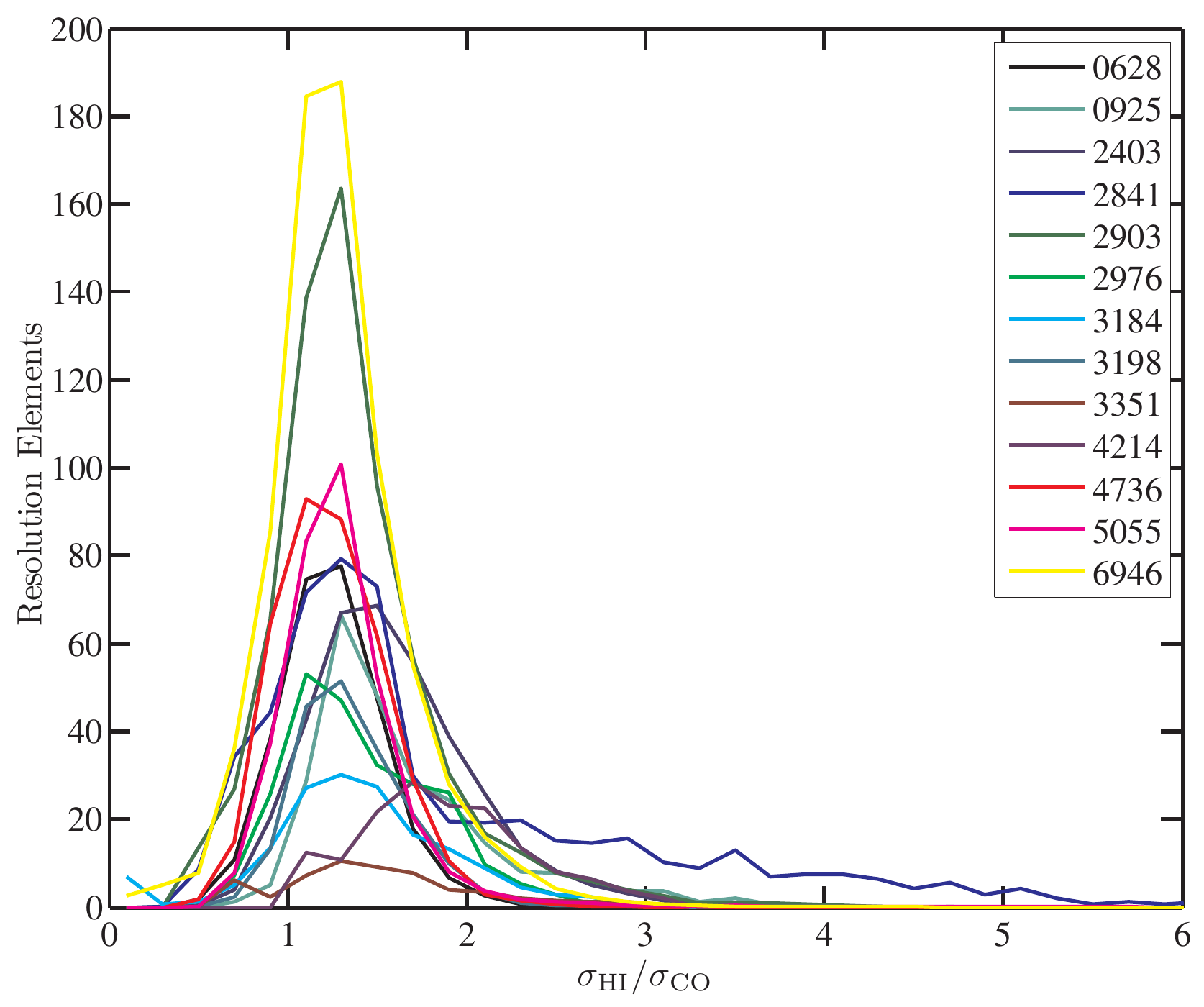}
\caption{Distributions of the dispersion ratios {\drat} of the
  galaxies for individual resolution elements.  The y-axis is in
  resolution elements (resolution elements = [number of
    pixels]/[number of pixels per single resolution element]).  Values
  for NGC 925, NGC 2841, NGC 2903, NGC 3198, NGC 3351, NGC 4214 and
  NGC 4736 are multiplied by a factor of 5 for better comparison with
  the other galaxies.}
\label{fig:DispRat}
\end{figure}

We constructed histograms of the {\ddif} values using bins of $1$
\kph.  For the {\drat} data we made histograms using bins of size
$0.2$.  Figure \ref{fig:DispDiff} shows the {\ddif} distributions for
each of the galaxies. Figure \ref{fig:DispRat} shows the {\drat}
distributions.  These all have Gaussian shapes and are symmetric and
well-sampled.  We therefore fitted Gaussians to the distributions.
The fitted mean dispersion difference and ratios are shown in
Table~\ref{tab:means}.  The CO distribution in NGC 925 and NGC 4214 only
encompasses a few resolution elements and care should be taken when
interpreting their fits.  The mean {\ddif} value is $-3.3 \pm 1.2$
\kph; the mean {\drat} is $1.4 \pm 0.2$ \kph.  A summary of the mean,
modes, and medians of the dispersion values is shown in Table
\ref{tab:dispval}.

{\citet{cal13}} quantified the effect of beam smearing in our galaxies
by simulating spiral galaxies with $30^{\circ}$, $60^{\circ}$, and
$80^{\circ}$ inclinations.  They found that beam smearing is greatest
in the central regions of galaxies and in highly inclined galaxies.
The observed dispersion can be increased by at most a factor of 1.2 at
$0.2$\,$r_{25}$ for galaxies with $30^{\circ}$ inclination, 1.5 for
$60^{\circ}$, and 1.8 for $80^{\circ}$, with these factors decreasing
quickly toward unity at larger radii.  They therefore used an
$0.2$\,$r_{25}$ radial cutoff for their analysis.  Even though
Figures {\ref{fig:HIDisp}} and {\ref{fig:CODisp}} and Table
{\ref{tab:comp}} show that including the inner pixels does not greatly
affect our conclusions, in our radial analysis and distribution width
analysis we only use pixels with radii greater than $0.2$\,$r_{25}$
which are therefore not affected by beam smearing.

 Our {\dhi} values are in agreement with {\citet{ler08}} ({\dhi}\,$ =
11 \pm 3$\,\kph), {\citet{ianja12}} ({\dhi}\,$ = 12.5 \pm
3.5$\,\kph) and {\citet{cal13}} ({\dhi}\,$ = 11.9 \pm 3.1$\,\kph)
who all analyzed the THINGS galaxies.    

\begin{table}
\begin{center}
\caption{The Mean, Mode and Median Dispersion Values \label{tab:dispval}}
\begin{tabular}{lrrrrr}
\tableline
\tableline
& \multicolumn{1}{c}{\dco} & \multicolumn{1}{c}{\dhi} & \multicolumn{1}{c}{\ddif} & \multicolumn{1}{c}{\drat}
\\
& \multicolumn{1}{c}{(\kph)} &  \multicolumn{1}{c}{(\kph)} &  \multicolumn{1}{c}{(\kph)} & 
\\
\\
\tableline
mode & 7.3 $\pm$ 1.7 & 11.7 $\pm$ 2.3 & $-$3.4 $\pm$ 1.4 &  1.3 $\pm$ 0.2
\\
mean & 10.5 $\pm$ 3.6 & 14.1 $\pm$ 4.3 & $-$3.3 $\pm$ 1.4 & 1.4 $\pm$ 0.2
\\
median & 9.5 $\pm$ 3.0 & 13.1 $\pm$ 3.0 & $-$3.3 $\pm$ 1.2 & 1.4 $\pm$ 0.2 
\\
fitted mean &  \multicolumn{1}{c}{-} &  \multicolumn{1}{c}{-} & $-$3.3 $\pm$ 1.2 & 1.4 $\pm$ 0.2 
\\
\tableline

\end{tabular}
\tablecomments{The values were calculated by taking the average of the
  modes, means, medians, and fitted means calculated for each galaxy in
  our sample. NGC 2841, NGC 2903, NGC 3198 and NGC 3351 were not
  included due their high inclinations and/or very
  asymmetric dispersion distributions.  The fitted mean was not
  calculated for the {\hi} and CO distributions because Gaussian
  profiles were not fitted to these distributions.}

\end{center}
\end{table}

\begin{figure*}[htp]
\centering
\includegraphics[width=0.8\textwidth]{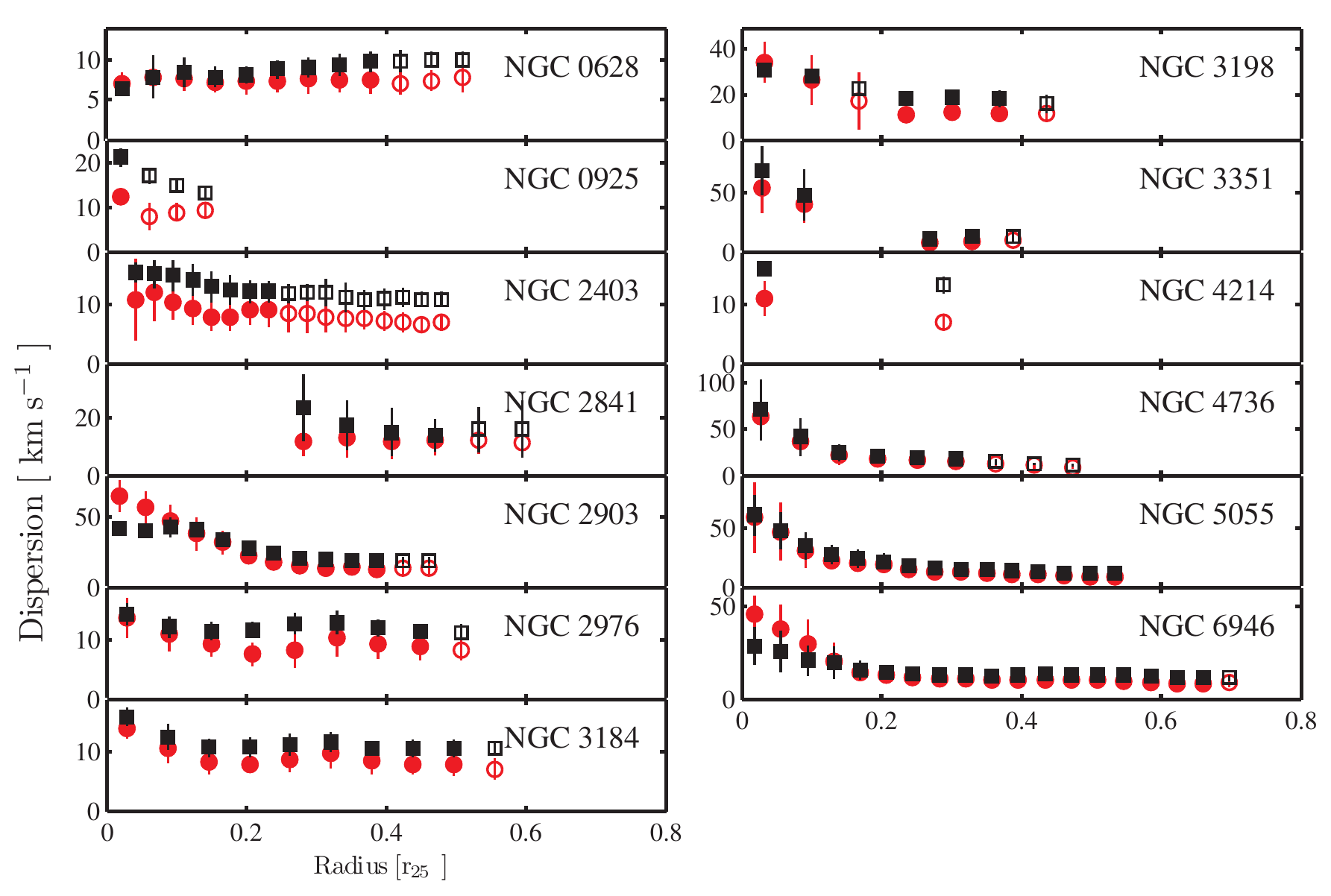}
\caption{The azimuthally averaged {\dco} (red circles) and {\dhi}
  (black squares) values plotted versus radius in each of the
  galaxies.  The radius is in units of $r_{25}$.  Azimuthal averages
  were taken over $13''$ annuli.  Data from annuli where the CO has a
  filling factor of more than $10$\% are plotted as open
  symbols. Annuli where the filling factor is more than $25$\% are
  plotted as filled symbols.  The error bars represent the standard
  deviation of the dispersion value in each annulus.}
\label{fig:COHIrad}
\end{figure*}

\begin{figure*}[htp]
\centering
\includegraphics[width=0.8\textwidth]{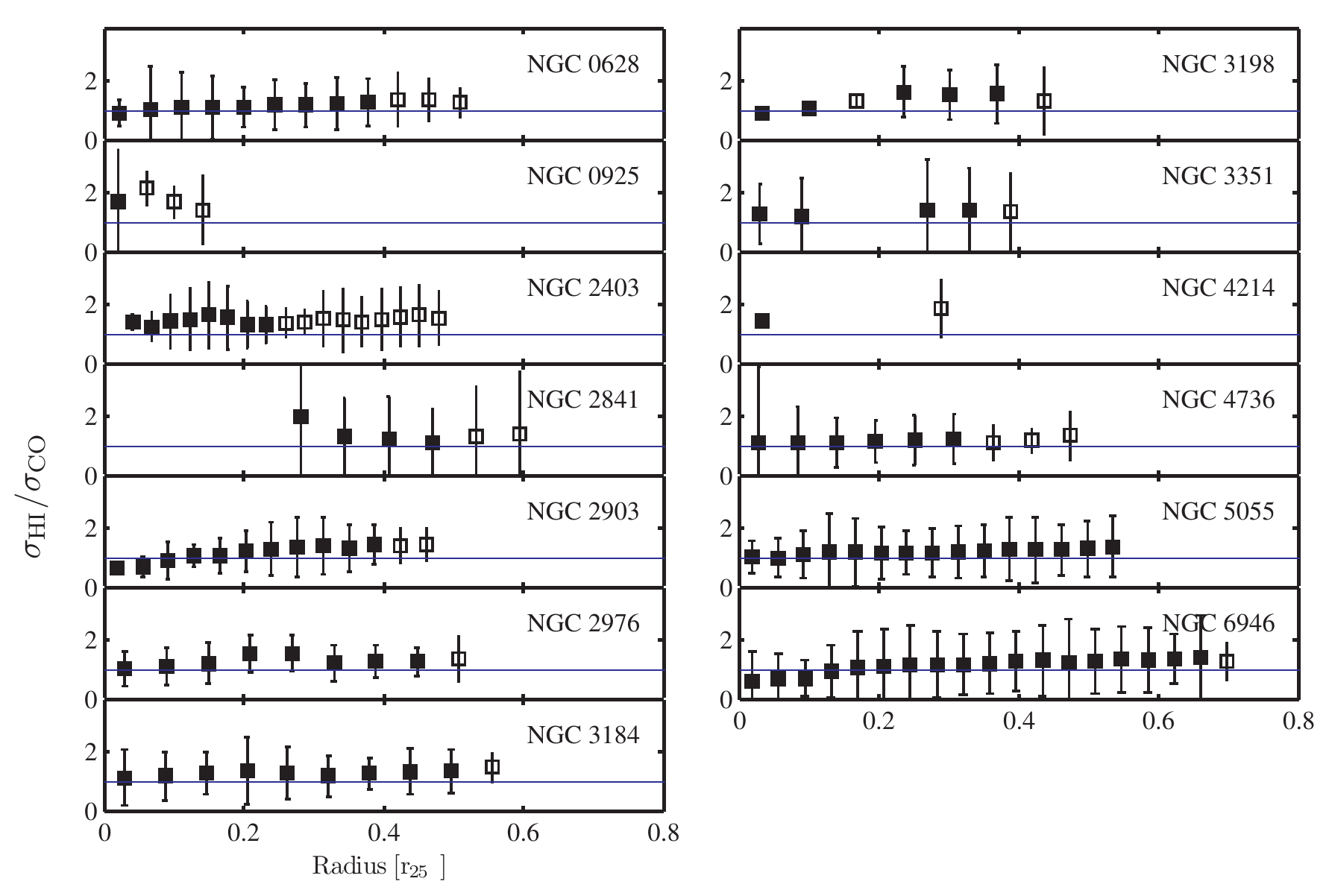}
\caption{The {\drat} ratio of the azimuthally averaged dispersions
  plotted versus radius in each of the galaxies.  The radius is in
  units of $r_{25}$.  Data from annuli where the CO has a filling
  factor of more than $10$\% are plotted as open symbols; data where
  the filling factor is more than $25$\% are plotted as filled
  symbols.  The horizontal lines indicate \dhi $=$ \dco.}
\label{fig:DRrad}
\end{figure*}

We also studied radial trends of the dispersions. The radial {\dhi},
{\dco} and {\drat} distributions are shown in Figs.\ \ref{fig:COHIrad}
and \ref{fig:DRrad}.  These plots were made using annuli where the
filling factors were higher than $10$\% and $25$\%, respectively.  In
our analysis we use the results from the 10\% annuli. Comparison with
the 25\% annuli shows that this choice of filling factor has little
effect on our results.  The {\drat} values in this analysis were
calculated by azimuthally averaging the {\drat} maps. \\

Figure \ref{fig:COHIrad} shows {\dco} and {\dhi} decreasing with
radius for most of the galaxies.  These also flatten off at larger
radii.  This behaviour was already seen in {\hi} by {\citet{tamb09}}.  
The {\drat} values remain roughly constant for most of the
radial range covered; this can be seen in Fig.\ \ref{fig:DRrad}.

\begin{table}
\begin{center}
\caption{Dispersion Values in \kph\ \label{tab:comp} for $r > 0.2 r_{25}$}
\begin{tabular}{l r r r r}
\tableline
\tableline
 & \multicolumn{1}{c}{$4$\,$S$} & \multicolumn{1}{c}{$8$\,$S$} & \multicolumn{1}{c}{Stacking} 
\\ 
 & \multicolumn{1}{c}{1} & \multicolumn{1}{c}{2} & \multicolumn{1}{c}{3} 
\\ 
\tableline

{\dco} mean & 9.3 $\pm$ 2.1 & 8.9 $\pm$ 2.1 & 12.8 $\pm$ 3.9
\\
{\dco} median & 8.6 $\pm$ 1.8 & 8.4 $\pm$ 2.0 & 12.0 $\pm$ 3.9 
\\
{\dhi} mean & 12.7 $\pm$ 2.1 & 12.3 $\pm$ 2.3 & 12.7 $\pm$ 3.1
\\
{\dhi} median & 12.2 $\pm$ 1.9 & 11.9 $\pm$ 2.1 & 11.9 $\pm$ 3.1 
\\
{\drat} mean & 1.5 $\pm$ 0.2 & 1.5 $\pm$ 0.3 & 1.0 $\pm$ 0.2 
\\
\tableline

\end{tabular}
\tablecomments{ The mean and median dispersion values determined for
  pixels at $r >0.2 r_{25}$.  Column 1: Using $4S$
  noise cutoff. Column 2: Using $8S$ noise cutoff. Column 3:
  Values from stacking analysis from {\citet{cal13}}.  }
\end{center}
\end{table}

\section{Comparison with Stacking Results}
\label{sec:discussion}

We now compare our results with the stacking analysis by
\citet{cal13}.  In that study, { which used the same data sets as
  the ones used here,} the stacking procedure included all possible CO
profiles, i.e., there was no rejection based on CO peak-flux and all
positions where an H{\,{\sc i}} velocity was available for use in
stacking the CO profile were used.  The only profiles excluded from
their analysis were those at radii less than $0.2 r_{25}$.  The mean
and median dispersion values they found are shown in Table
\ref{tab:comp}.  These {\dco} values are higher than our mean and
median values.

To check whether this difference is not caused by the higher
  uncertainties associated with low peak-flux profiles, we rederived
  our values for a number of different noise cutoffs between $4S$
and $8S$.  Our mean and median dispersion values for data with the
central $0.2 r_{25}$ pixels removed and using $4S$ and $8S$ noise
cutoffs are shown in Table \ref{tab:comp}.  Our pixel-by-pixel {\dco}
values are lower irrespective of which noise cutoff we use.  Our
{\dhi} values remain similar to the \citet{cal13} values. 
 
Due to the noise cutoff used here, our analysis does not probe the
low peak-flux regime that the stacking analysis in \citet{cal13} is
sensitive to.  It is therefore possible that the difference found
  in dispersion values could be caused by profiles with a peak flux
  lower than $4S$ having systematically higher velocity
  dispersions. It is, however, difficult to directly and accurately
  measure the individual velocity dispersions of these low peak-flux
  profiles.

\begin{figure*}[htp]
\centering
\includegraphics[width=0.8\textwidth]{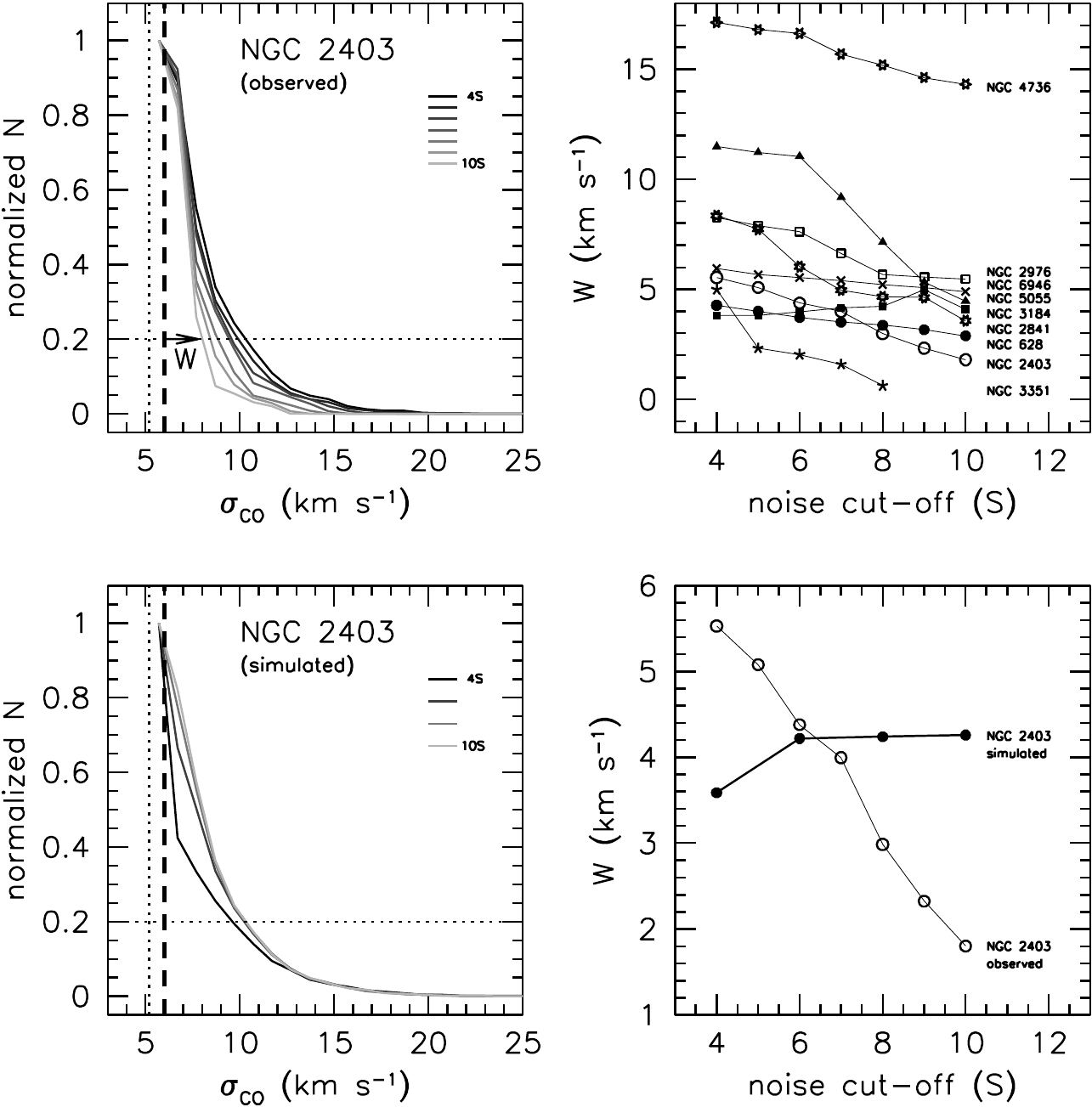}
\caption{{\it Top Left:} normalized distributions of $\sigma_{\rm CO}$
  for NGC 2403 for different values of the noise cutoff. The black
  curve indicates a $4S$ cutoff. From black to light gray, the noise
  cutoff increases in steps of $S$ with the light-gray curve
  indicating a $10S$ cutoff. It is clear that the curves become narrower
  with increasing noise cutoff. The thick dashed vertical line
  indicates the mode of the distribution, the dotted vertical line the
  velocity resolution cutoff. The horizontal dotted line indicates
  the 20 percent level at which the width with respect to the mode is
  measured. The width $W$ is indicated for the $10S$ curve by the arrow
  labelled ``W''. {\it Bottom Left:} normalized distributions of the
  simulated NGC 2403 data for different peak-flux values. Velocity
  profiles were simulated with random noise, varying input amplitudes,
  and input dispersions drawn from NGC 2403's $4S$ {\dco}
  distribution.  The profiles were fitted in the same manner as the
  observed data and the histograms of the fitted dispersions are
  plotted.  {\it Top Right:} the 20 percent width $W$ as defined in
  the left panel plotted against the noise cutoff in units of
  $S$. Galaxies are labeled to the right of their corresponding
  curve. Note that for NGC 3351 only values up to $8S$ could be
  measured. {\it Bottom Right:} the $W$ values of the observed NGC
  2403 {\dco} distributions and the simulated distributions are
  plotted against the modeled peak-flux value. }
\label{fig:width}
\end{figure*}

We therefore evaluated the impact of the low peak-flux spectra by
creating histograms of the velocity dispersion values for different
noise cutoff values. If lower peak-flux profiles do indeed have
  higher velocity dispersions, then we would expect the fraction of
  high-dispersion profiles to decrease with increasing noise cutoff.
In other words, the prominence of any high-dispersion tail in the
histogram should decrease.  For this analysis we used data from pixels
with radii greater than $0.2$\,$r_{25}$.  An example is shown in the
top left panel of Fig.\ \ref{fig:width}.  Here we show the normalized
$\sigma_{\rm CO}$ distributions for NGC 2403 derived using various
noise cutoffs between $4S$ and $10S$. It is clear that the
distribution becomes more narrow with increasing cutoff value. We
quantify this with the histogram half-width at 20 percent of the
maximum, where the half-width is measured in the direction of higher
dispersions, with respect to the value of the histogram maximum (or
the mode), as indicated in the left panel of Fig.\ \ref{fig:width}.
We have repeated this analysis for all our sample galaxies, except for
NGC 925, NGC 4214 and NGC 3198, where CO emission is faint and limited
in extent, and NGC 2903, where the dispersion values are dominated by
streaming motions along the bar. We also excluded the central part of
NGC 3351 which is dominated by a compact bar.

\begin{figure}
\centering
\includegraphics[width=0.45\textwidth]{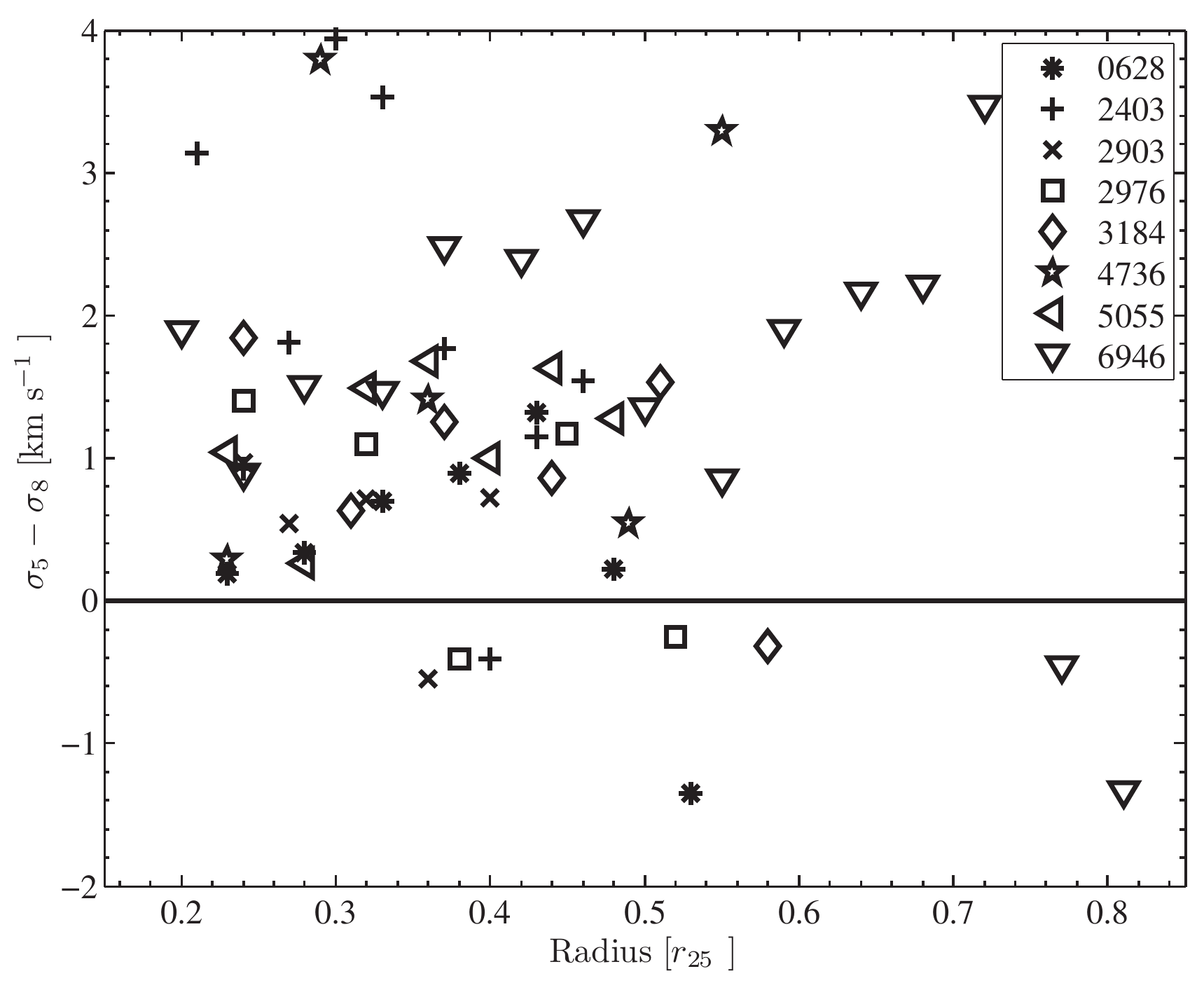}
\caption{Differences between the stacked mean {\dco} dispersion
  calculated for profiles with peak flux $> 8S$ (${\sigma}_{8}$) and
  $>5S$ (${\sigma}_{5}$) plotted against radius.}
\label{fig:stack2}
\end{figure}

The top-right panel of Fig.~\ref{fig:width} shows the values of the
measured widths as a function of noise cutoff. For the majority of
the galaxies shown there, the width becomes narrower toward higher
cutoff values and the high-dispersion tail less prominent. We thus
find that the fraction of high-dispersion spectra does indeed decrease
with increasing peak flux and it is therefore likely that the low peak-flux
profiles included in the stacking analysis (but excluded in ours) have
systematically higher velocity dispersions. 

This implies that repeating the stacking analysis of \citet{cal13}
with a \emph{higher} noise cutoff should give \emph{lower} dispersion
values than their original analysis.  We therefore performed a stacking
analysis on our data using various cutoffs. As an example, in Figure
\ref{fig:stack2} we show, as a function of radius, the difference
between the $5S$ and the $8S$ stacked dispersions. The average
difference between the dispersions is $\sim 1.5$ \kph, independent of
radius, with the $5S$ values systematically higher than the $8S$ ones.

To test whether the trend in velocity dispersion we found is not
caused by the increasing importance of the noise toward lower peak
fluxes (which might be expected to broaden profiles), we repeat our
analysis using simulated profiles.  We created Gaussian profiles with
peak values between $4S$ and $10S$, and with dispersions chosen with a
probability distribution equal to the observed {\dco} distribution of
NGC 2403, as shown in Fig.~\ref{fig:CODisp} (for pixels with radii
greater than $0.2$\,$r_{25}$).  We explicitly assume that the velocity
dispersion is independent of the peak flux.  We compare the dispersion
distributions found at various peak-flux levels in Figure
{\ref{fig:width}} (bottom left). The change in these distributions is
very different from that as observed for NGC 2403, as shown in the
bottom right panel in Figure {\ref{fig:width}}.  The higher
  velocity dispersions measured at low peak flux are therefore not due
  to noise affecting the profiles, but due to an increase of the
  velocity dispersion toward lower peak fluxes.

A similar anti-correlation, but for {\hi} rather than CO, was found
for a number of dwarf galaxies by \citet{hunter01,hunter11}.
\citet{hunter11} suggest that this anti-correlation is roughly
consistent with a uniform pressure throughout these galaxies, as also
found in simulations of magneto-rotational instabilities by
\citet{piontek05}.
  
Returning to the observed CO velocity dispersions, \citet{cal13} note
that the higher dispersion values that are found in the stacked
profiles can be explained with a diffuse, extended molecular gas
component that pervades our galaxies in addition to the molecular gas
in GMCs in the thin, ``cold'' CO disk.  Our pixel-by-pixel analysis is
limited to pixels with bright CO emission, which is dominated by the
GMCs.  These velocity profiles are narrower than those dominated by
emission from the diffuse CO disk.

These differences therefore are further evidence that a diffuse,
high-dispersion component of molecular gas is present in our galaxies
in addition to a thin molecular disk.  The diffuse component of
molecular disks may thus be a common feature in disk galaxies.

\section{Summary}

We have measured the velocity dispersions in individual \hi\ and CO
profiles of a number of THINGS disk galaxies. We find an \hi\ velocity
dispersion of {\dhi}$ = 11.7 \pm 2.3$ \kph. The corresponding CO value
is {\dco}$ = 7.3 \pm 1.7$ \kph. The ratio between these two
dispersions is {\drat}$ = 1.4 \pm 0.2$ and is not correlated with
radius.

In a previous study using the same data, \citet{cal13}, by stacking
individual velocity profiles, found a systematically higher CO
velocity dispersion and a ratio {\drat}$ = 1.0 \pm 0.2$.  This
difference can be explained if low peak-flux CO profiles have a
systematically higher velocity dispersion than high-peak flux
profiles.  Our pixel-by-pixel analysis preferrentially selects the
bright, high peak-flux CO profiles, in contrast with the stacking
analysis which also includes large numbers of low peak-flux CO
profiles.

The relation of \dco\ decreasing with increasing profile amplitude is
consistent with a picture where the bright CO regions (preferentially
selected in studies of individual profiles) are dominated by
narrow-line GMCs, with a more diffuse, higher dispersion component
(more efficiently detected in stacking analyses) becoming more
prominent toward lower intensities.  A pixel-by-pixel analysis is
therefore a good way to study the thin molecular disk component where
GMCs dominate the emission. In turn, stacking analyses are more
sensitive to the diffuse, high-dispersion extended molecular disk
component.

Our results thus provide further evidence for the suggestion presented
in {\citet{cal13}} that many disk galaxies have an extended, diffuse
molecular disk component in addition to a thin, GMC-dominated,
molecular disk.

\acknowledgments
We gratefully thank the anonymous referee for all the comments and suggestions 
that improved the content of the paper.  K.M.M. gratefully acknowledges support 
from the Square Kilometre Array South Africa (SKA-SA) and the National 
Astrophysics and Space Science Program (NASSP). K.M.M. would like to thank 
S. Schutte for the useful discussions during the preparation of the manuscript. 
The work of W.J.G.d.B. was supported by the European
Commission (grant FP7-PEOPLE-2012-CIG \#333939).  R.I. acknowledges
funding from the National Research Foundation (NRF grant number
MWA1203150687) and the University of South Africa (UNISA) postdoctoral
grant.

Facilities: \facility{IRAM 30m}, \facility{VLA}.

\appendix
\section{Comparing {\rm \hi} velocity dispersions and second moments}

In a previous study, \citet{tamb09} determined the second moments of
the {\hi} profiles of the THINGS galaxies as an estimate for the
velocity dispersions. These second moments were measured as a function
of radius over the full extent of the \hi\ disk.

To gauge how well the second-moment values match the Gaussian
dispersion values \dhi, we derive both parameters for our
\hi\ profiles, as also measured over the entire radial range and full
area of the \hi\ disk, i.e., also including regions without CO
emission.

\begin{figure*}
\centering
\includegraphics[width=0.8\textwidth]{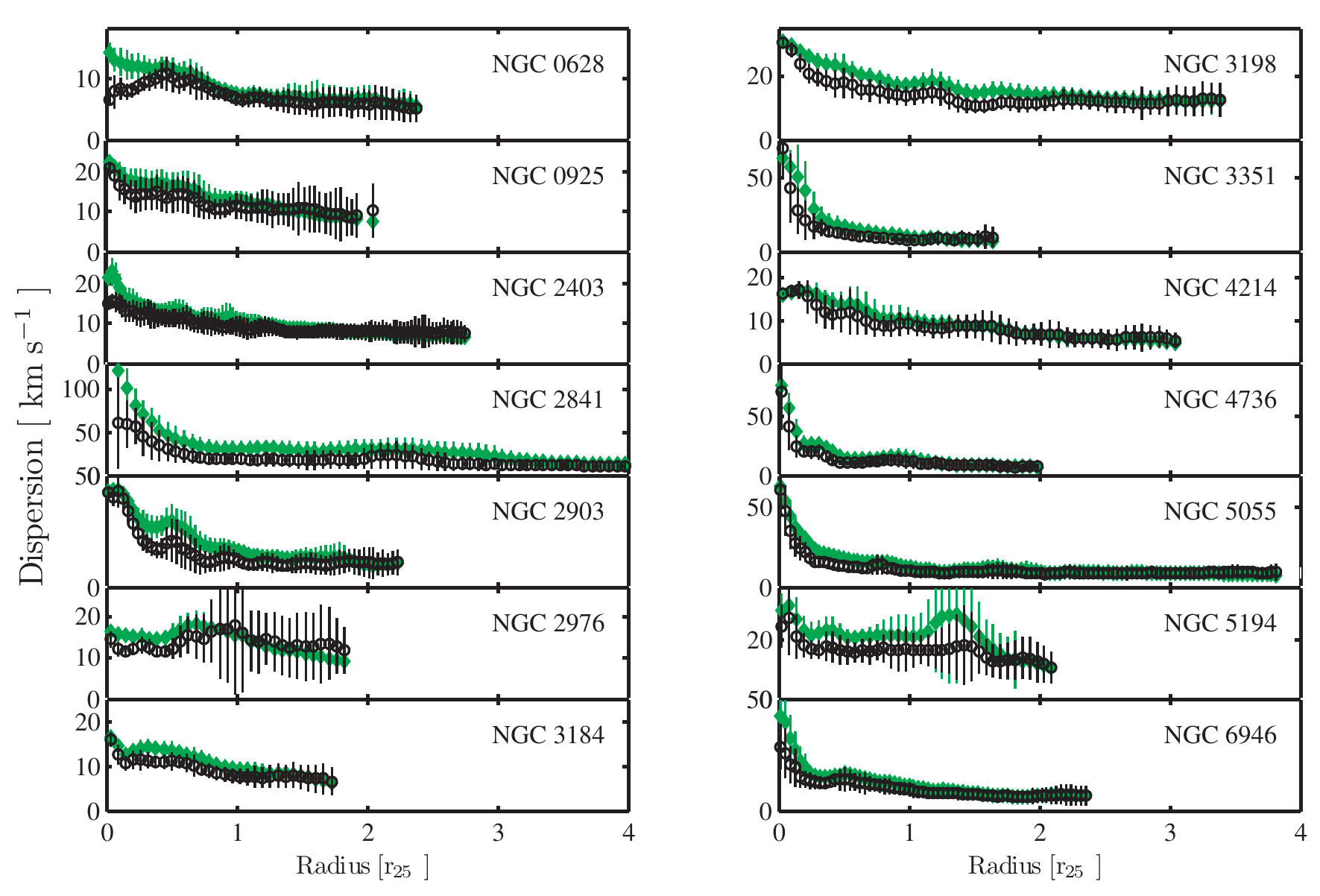}
\caption{Azimuthally averaged {\dhi} values plotted versus radius
  in each of the galaxies.  The radius is in units of $r_{25}$.
  Azimuthal averages were taken over $13''$.  The error
  bars represent the standard deviation of the dispersion values in
  each annulus.  The black open circles are Gaussian fitted dispersions and
  the green filled diamonds are second-moment dispersions. Only annuli with
  filling factors above $10$\% are shown.}
\label{fig:HIallrad}
\end{figure*}

Figure \ref{fig:HIallrad} shows that
there are some slight differences in the second-moment values and
Gaussian fitted dispersions.  For most galaxies the largest
differences between second-moment values and Gaussian fitted
dispersions are found in the inner regions of galaxies, with
second-moment values being larger than the Gaussian fitted
dispersions.  The inner regions of the galaxies have more non-Gaussian
profiles than the outer regions. This shows that the second moment is more
sensitive to non-Gaussianities than profile fits and in these cases
should be interpreted with care.  

We note that the {\hi} dispersions associated with the CO disk (the
inner star-forming disk) are higher than the dispersion as measured
over the entire {\hi} disk (which includes the outer parts of the
galaxy where there is no detectable CO).  This is can be explained by
the higher star formation rate in the inner disk compared to the outer
disk. A further discussion of this is, however, beyond the scope of
this paper.

\end{document}